\documentclass[twocolumn,prb,amsmath,amssymb]{revtex4}
\usepackage{graphicx,bm}

\def\bp{{\mathbf p}}
\def\bq{{\mathbf q}}
\def\bj{{\mathbf j}}
\def\bv{{\mathbf v}}

\def\bF{{\mathbf F}}

\def\bnab{{\bm \nabla}}

\newcommand{\de}{\delta}
\newcommand{\D}{\Delta}
\newcommand{\e}{\epsilon}

\newcommand{\Lm}{\Lambda}
\newcommand{\s}{\sigma}
\newcommand{\del}{\nabla}
\newcommand{\p}{\partial}

\newcommand{\bsdel}{\boldsymbol{\nabla}}

\newcommand{\mbj}{\mathbf{j}}

\newcommand{\mbp}{\mathbf{p}}
\newcommand{\mbps}{{\mathbf{p}\sigma}}

\newcommand{\mbq}{\mathbf{q}}

\newcommand{\mbr}{\mathbf{r}}

\newcommand{\mbv}{\mathbf{v}}

\newcommand{\bsphi}[1]{\boldsymbol{\phi}}
\newcommand{\mbf}[1]{\mathbf{#1}}

\newcommand{\bfhat}[1]{\hat{\mathbf{{#1}}}}

\newcommand{\til}[1]{\tilde{#1}}
\newcommand{\pfrac}[2]{\left(\frac{#1}{#2}\right)}
\newcommand{\bs}[1]{\boldsymbol{#1}}
\newcommand{\nn}{\nonumber\\}

\newcommand{\ben}{\begin{equation}}
\newcommand{\een}{\end{equation}}

\begin{document}

\title{Spin-Seebeck effect in a strongly interacting Fermi gas}
\author{C. H. Wong, H.T.C. Stoof and R.A. Duine}
\affiliation{Institute for Theoretical Physics, Utrecht University, Leuvenlaan 4, 3584 CE Utrecht, The Netherlands}
\date{\today}
\begin{abstract}
We study the spin-Seebeck effect in a strongly interacting, two-component Fermi gas and propose an experiment to measure this effect by relatively displacing  spin up and spin down atomic clouds in a trap using spin-dependent temperature gradients.   We compute the spin-Seebeck coefficient and related spin-heat transport coefficients as functions of temperature and interaction strength.  We find that when the inter-spin scattering length becomes larger than the Fermi wavelength, the spin-Seebeck coefficient changes sign as a function of temperature, and hence so does the direction of the spin-separation.  We compute this zero-crossing temperature as a function of interaction strength and in particular in the unitary limit for the inter-spin scattering.   
\end{abstract}
\maketitle
\textit{Introduction}.--
Spin caloritronics, the study of coupled spin and heat transport, is a rapidly developing subfield of spintronics [\onlinecite{bauerSCC10}].   In particular, the spin-dependent generalization of the Seebeck effect, called the spin-Seebeck effect, has been intensively studied in the solid-state environment [\onlinecite{hatamiSCC10}].   Recently, there has been broad interest in exploring spintronic phenomena in cold atomic systems [\onlinecite{linNAT11,duinePRL09,wongPRL12}].   Spin transport in a strongly interacting, two component Fermi gas was investigated experimentally in Ref.~[\onlinecite{sommerNAT11}].   It is the purpose of this Letter to study the associated heat transport, i.e., thermo-spin effects, in a similar setting.

\begin{figure}[t]
\begin{center}
\includegraphics[width=.7\linewidth]{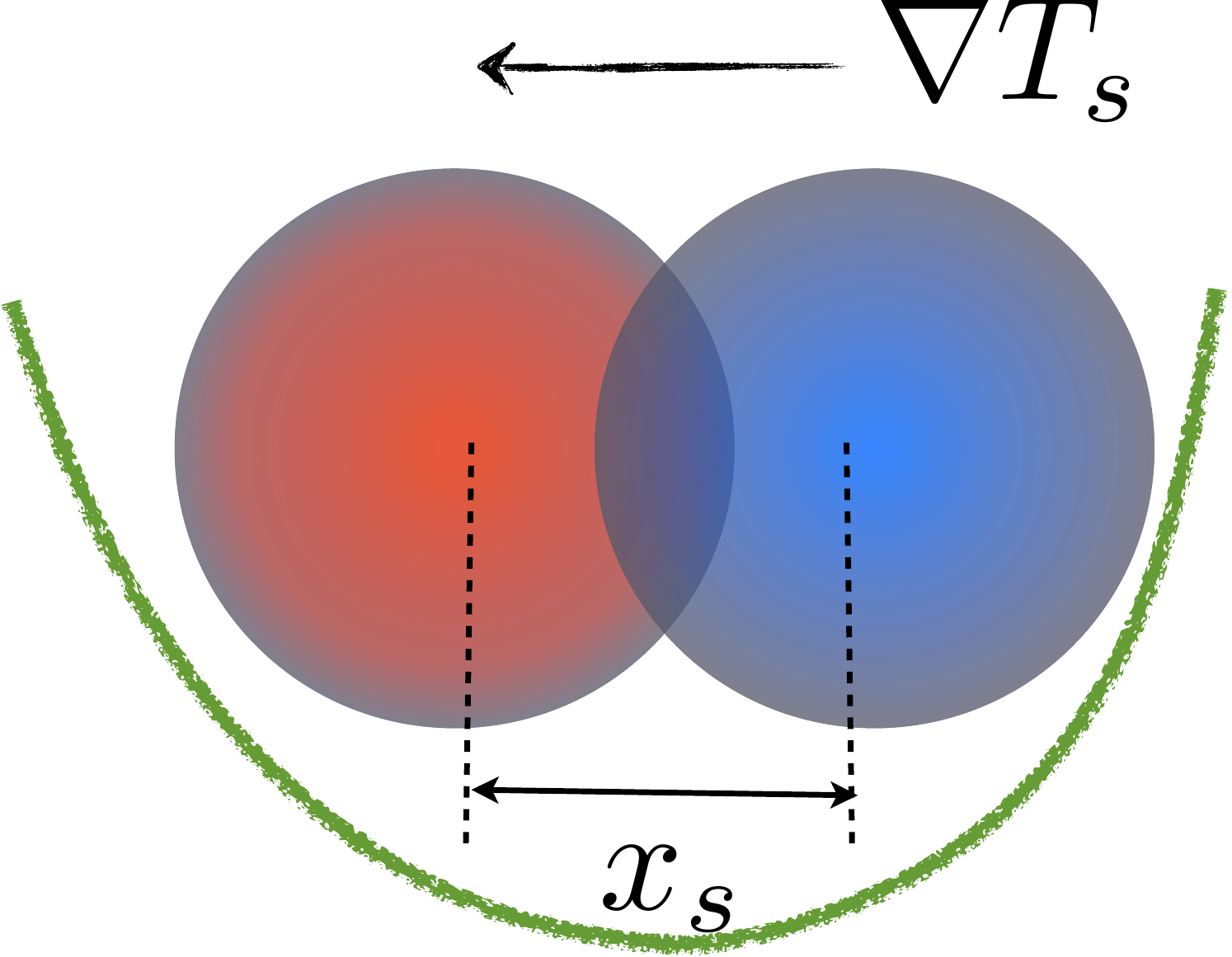}
\caption{Spin up and spin down atomic clouds are spatially relatively displaced in the presence of a spin temperature gradient.  The distance between the center of mass of the different spin components, denoted by $x_s$, is proportional to the spin-Seebeck coefficient.}
\label{cartoon}
\end{center}
\end{figure}

In the ordinary Seebeck effect in metals, an electrochemical potential gradient is generated by applying a temperature gradient.  Similarly, for a gas with two spin states, the spin-Seebeck coefficient $S_s$ determines the spin chemical potential $\mu_s$ generated by a spin temperature gradient $\del T_s$ through the relation $\del\mu_s=S_s\del T_s$, where $\mu_s\equiv\mu_+-\mu_-$ and $T_s\equiv T_+-T_-$, $\mu_\s$ and $T_\s$ being the spin-dependent chemical potential and temperatures of the spin $\s$ atoms, respectively, and we label spin components by $+$ and $-$.   To measure the spin-Seebeck coefficient, we propose relatively displacing the center of mass of spin up and spin down atom clouds in a harmonic trap by applying a spin-dependent temperature gradient, for example by selectively heating one spin component with a laser, as illustrated in Fig.~\ref{cartoon}.  The locations $x_\pm$ of the center of mass of the spin up and down atoms are shifted to the minimum of $\mu_\pm+V$, where  $V$ trapping potential, resulting in a spin separation $x_s=x_+-x_-={S_s\nabla  T_s}/{m \omega ^2}$,
where $m$ is the mass of the atoms, $\omega$ is the trap frequency in the direction of the temperature gradients.
For an order of magnitude estimate, we take $S_s\simeq.01k_B$, as verified below.  For $\nabla T_s= 10^{-5}\,\text{K/cm}$ [\onlinecite{meppelinkPRL09}], $\omega=2\pi\times1.46$, we find $x_s\simeq1$ mm, which is well within experimental resolution.


\begin{figure}[t]
\begin{center}
\includegraphics[width=\linewidth]{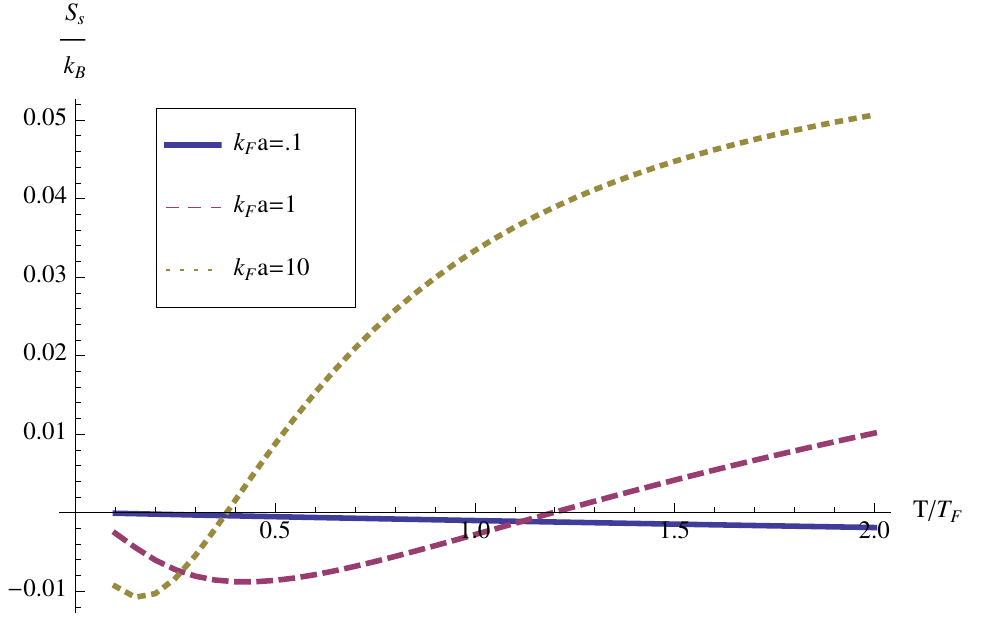}
\caption{The Seebeck coefficient plotted as a function of the reduced temperature $T/T_F$, where $T_F$ is the Fermi temperature, for  values of $k_Fa$ representing weak ($k_Fa=.1$) and strong ($k_Fa=1$) coupling, and approaching the unitary limit ($k_Fa=10$).
}
\label{seebeckfig}
\end{center}
\end{figure}

\begin{figure}[t]
\begin{center}
\begin{tabular}{cc}
\includegraphics[width=0.5\linewidth]{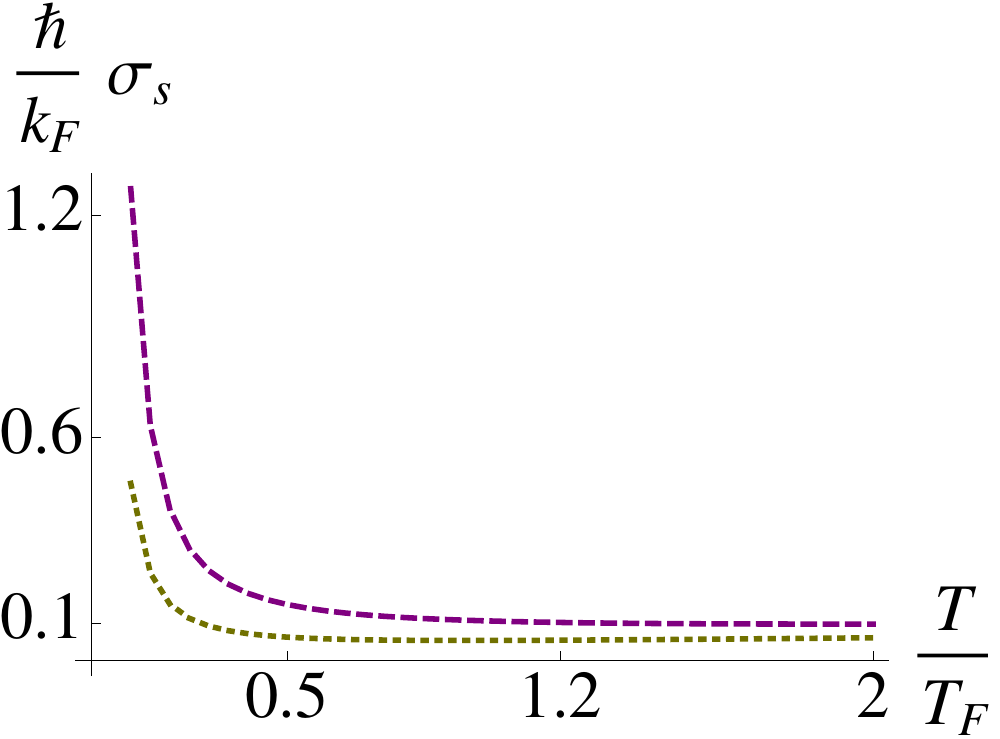}&
\includegraphics[width=0.5\linewidth]{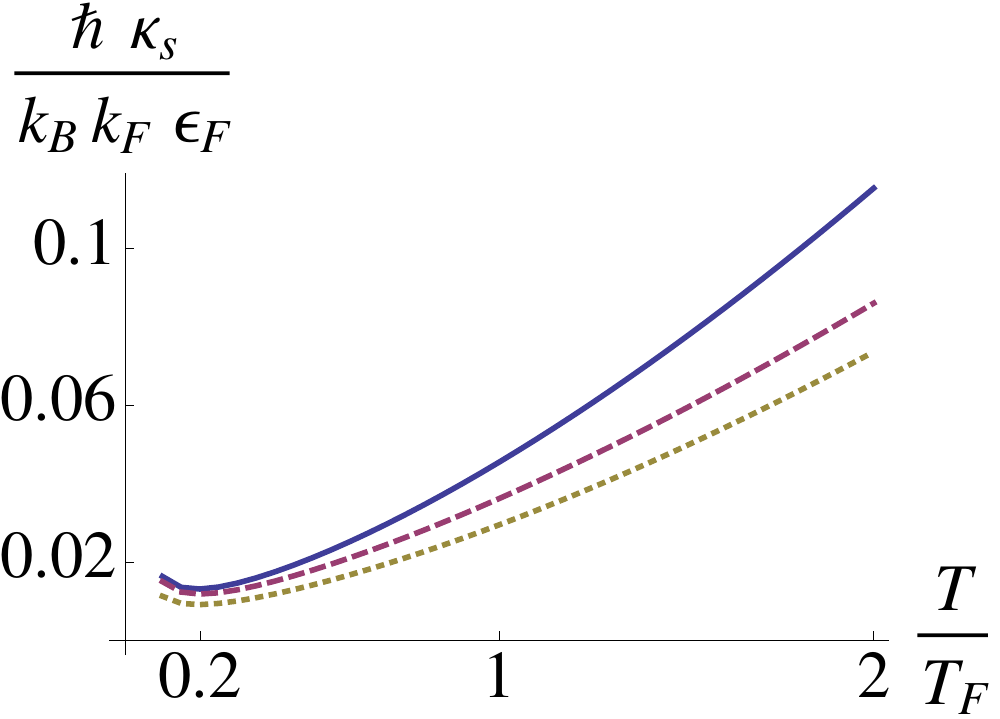}\\
(a)&(b)\\
\includegraphics[width=0.5\linewidth]{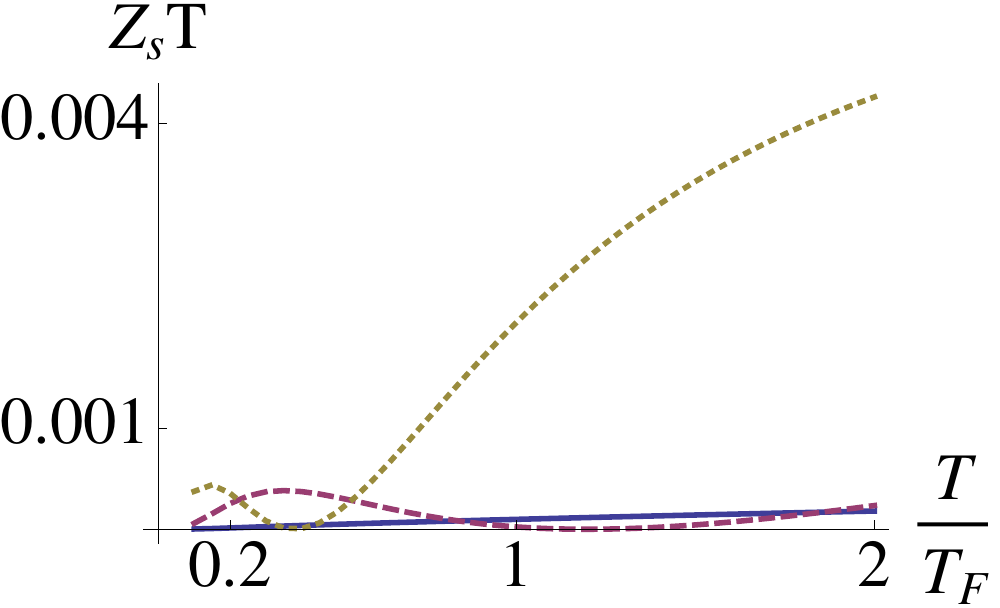}&
\includegraphics[width=0.5\linewidth]{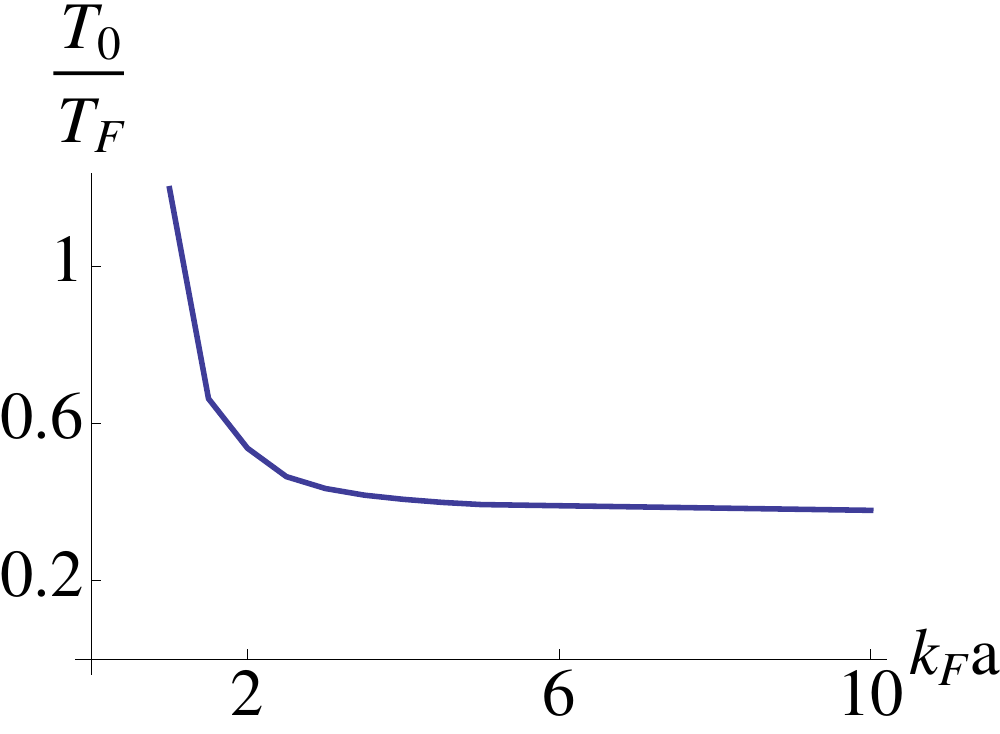}\\
(c)&(d)
\end{tabular}
\caption{(a) A plot of the spin conductivity normalized as $\hbar \Lm \s_s$, (b) the spin heat conductivity normalized as $\hbar \Lm \kappa_s/k_B\e_F$, and  (c) {the figure of merit $Z_sT$, for $k_Fa=.1$ (thick blue), $k_Fa=1$ (dashed purple), $k_Fa=10$ (dotted yellow).} (d)  A plot of the temperatures $T_0$ relative to $T_F$  where $S_s=0$ as a function of $k_Fa$.}
\label{spincond}
\end{center}
\end{figure}

We have computed the spin-Seebeck coefficient for a two-component Fermi gas, plotted in Fig.~\ref{seebeckfig}, as a function of temperature and for several values of the interaction strength $k_Fa$, where $k_F$ is the Fermi wave vector and $a$ is the inter-spin scattering length.  As seen from the figure, for weak interactions ($k_Fa<1$),  $S_s$ is small and negative, while for strong interactions ($k_Fa\geq1$),  $S_s$ is larger and its sign changes as a function of temperature.   In terms of the experiment mentioned above, this means that the spin displacement changes direction as a function of temperature, which is an interesting qualitative effect.    We also plot the zeros of $S_s$ as a function of $k_Fa$ in Fig. \ref{spincond}d.   The temperature of the zero-crossing reaches a universal value $T_0$, in the unitary limit for the inter-spin scattering length $k_Fa\to\infty$,  We find $T_0\simeq.378\,T_F$ in our calculation, where $T_F$ is the Fermi temperature.   

The thermodynamic reciprocal of the spin-Seebeck effect is the spin-Peltier effect, in which a spin-dependent heat current  proportional to the spin-Seebeck coefficient is induced by a spin current. This effect will heat up spin-up and spin-down components differently and provides another way to measure $S_s$.  Furthermore, as discussed in Ref.~[\onlinecite{wongPRL12}], the spin-Seebeck effect contributes to the total dissipation so that $S_s$ can also be measured through the heating.  
 
 We note that the spin-Seebeck coefficient was calculated for a weakly interacting Bose gas in Ref.~[\onlinecite{wongPRL12}], but the Bose gas is unstable towards the formation of molecules for large scattering lengths, which makes the strongly interacting regime more difficult to realize experimentally. 


\textit{Phenomenology}.-- We are specifically interested in phenomena due to \emph{spindrag}, the transfer of momentum between different spins due to inter-spin scattering, which allows one to generated currents in one spin species by applying forces on the other, and we define a set of spin-heat transport coefficients which captures these effects as follows.  We consider a Fermi gas with two different spin states selected from a larger half-integer spin multiplet, which we will call ``spin up" ($+$) and ``spin down" ($-$), for equal spin up/down densities  $n_+=n_-\equiv n$, i.e.,  in the absence of spin polarization, and apply equal and opposite forces and temperature gradients for the two spin species, i.e., $\bF_+ = -\bF_- $ and $\bnab T_+ = -\bnab T_- $.  In linear response, the ensuing spin current and spin heat current defined by $\bj_{\rm s} = \bj_+ - \bj_-$, and $\bq_{\rm s} = \bq_+ - \bq_-$, respectively, are given by
\begin{align}
\left(\begin{array}{c}\mbj_s \\ \mbq_s\\\end{array}\right) 
=\s_s\left(\begin{array}{cc} 1& S_{ s} \\ 
TS_{ s} & {\kappa_s\over\s_s} (1+ Z_sT)
 \\ \end{array} \right) 
\left( \begin{array}{c} \mbf{F}_{ s} \\ -\bsdel T_{ s} \\\end{array} \right)\,,
\label{response}
\end{align}
where $\mbf{F}_s\equiv\mbf{F}_+ -\mbf{F}_-$ is the spin force, $T_s=T_+-T_-$ is the spin temperature, $T$ is the equilibrium temperature, $\sigma_s$ is the spin conductivity,  $\kappa_{\rm s}$ is the spin heat conductivity (at zero spin current), $Z_{\rm s} T = \sigma_{\rm s} S_{\rm s}^2 T /\kappa_{\rm s}$,  and Onsager reciprocity is explicitly included in the matrix above.    We note that $\mbf{F}_s$ is the thermodynamic force which includes forces coming from pressure gradients, i.e., $\mbf{F}_s=\mbf{f}^{\rm ext}_s-\bsdel p_s/n$, where $\mbf{f}^{\rm ext}_s$ is the external spin force, and $p_s=p_+-p_-$ is the difference in pressures of the spin up and down atoms.    These coefficients, computed with the Boltzmann equation described below, are plotted in Fig.~\ref{spincond} as functions of $T/T_F$ for several values of $k_Fa$.   

As is well known, the spin conductivity $\s_s$ rapidly increases at low temperatures due to Pauli blocking.  Our result for $\s_s$, plotted in Fig. \ref{spincond}a includes corrections due to spin-heat coupling, but they are negligibly small, so that one can safely take $\s_s=n\tau_s/m$ with $\tau_s$ the spindrag relaxation time [\onlinecite{trapavg}] measured in Ref.~[\onlinecite{sommerNAT11}] and calculated in  Ref.~[\onlinecite{bruunNJP11}].  The downturn of $S_s$ at low temperatures is a quantum mechanical effect that also occurs for bosons, where in contrast to fermions, the spin conductivity decreases sharply at low temperatures due to bosonic enhancement of scattering [\onlinecite{drielPRL10}].  The spin heat conductivity $\kappa_{\rm s}$  is plotted in Fig.~\ref{spincond}b, where it is seen to increase with increasing $T$.   The dimensionless figure of merit $Z_{\rm s} T$ (Fig.~\ref{spincond}c)  determines the thermodynamic efficiency of engines based on thermo-spin effects [\onlinecite{mahanSSP97}].

A spin-dependent temperature gradient can only be established when intra-spin scattering is much greater than inter-spin scattering.  For fermions, the first nonvanishing intra-spin scattering amplitude is $p$-wave.  To make this large, one can tune the $p$-wave scattering length by a Feshbach resonance [\onlinecite{regalPRL03}].  Taking the unitary limit for intra-spin scattering, the intra and inter-spin differential cross section are given by
\begin{align}
\frac{{d\sigma}_{++}}{{d\Omega}}&=\frac{{d\sigma}_{--}}{{d\Omega}}=\frac{9(\bfhat{p}_r\cdot\bfhat{p}_r')^2}{(p_r/\hbar )^2}\,,\nn
\frac{{d\sigma}_{+-}}{{d\Omega}}&=\frac{a^2}{1+(p_ra/\hbar )^2}\,,
\label{cross}
\end{align}
where $p_r=|\mbp_r|$ is the relative momentum of incoming particles with momenta  $\mbp_1$ and $\mbp_2$, defined by $\mbp_r=(\mbp_1-\mbp_2)/2$, the hat superscripts denotes unit vectors, and $a$ is the inter-spin $s$-wave scattering length.

 \textit{Calculation of transport coefficients}.--
Next, we present the computation of $S_s$ using the Boltzmann equation.  We parametrize the non-equilibrium, steady-state distribution by 
\ben
 n_ {\mbp\s}(\mbr)=f_ {\mbp\s}(\mbr)-\p_\e f^0_\mbp \phi_{\mbps}(\mbr),
 \een
where $f^0_\mbp=(\exp[(\e_\mbp-\mu)/k_BT]+1)^{-1}$ is the equilibrium Fermi distribution, $\mu$ is the chemical potential,    $\e_\mbp=\mbp^2/2m$, and $f_ {\mbps}(\mbr,t)=(\exp{[\e_\mbp-\mu_\s(\mbr))/k_BT_\s(\mbr)}]+1)^{-1}$  is the local equilibrium distribution, $\p_\e f^0_\mbp=-{f^0_\mbp(1-f^0_\mbp)/k_BT}$, and $\phi_\mbps$ is determined by solving the Boltzmann equation for the spin distribution $n_{\mbp s}=n_{\mbp+}-n_{\mbp-}$ in linear response,
\ben
\p_\e f^0_\mbp \left({\e_\mbp-w(T)\over k_BT}\right)\mbv_\mbp=\mbf{C}_\mbp[\bs{\phi}],
\label{bolt}
\een
where $w(T)=\mu+Ts$ is the enthalpy per particle and $s$ is the entropy per particle [\onlinecite{enthalpy}]. We defined $\phi_{\mbp s}\equiv k_B\bs{\phi}_\mbp\cdot(-{\bsdel T_s })$, and expressed the linearized collision integral in the Boltzmann equation for the spin distribution  as $(\p n_{\mbp s}/\p t)_{\rm coll}\equiv \mbf{C}_\mbp[\bs{\phi}_\mbp] \cdot(-\bsdel T_s)$.  The spin current is given by
 \begin{align}
 \mbf{j}_s&=-\int{d^3p\over(2\pi\hbar)^3}\p_\e f^0_\mbp\,\bv_\bp \phi_{\mbp s}.
 \label{current}
 \end{align}


We solve  Eq.~\eqref{bolt} using the method described in Ref.~[\onlinecite{wongPRL12}].  Applying the temperature gradient along the $x$-axis,  we parametrize the response by a power series,
 \ben
 {\phi}_{\mbp s}(a,T)=\left[b_0(a,T)+b_1(a,T)\pfrac{\e_\mbp}{k_BT}\right]p_x(-k_B\p_x T_s)\,.
 \label{phis}
 \een
The coefficients $b_0$ and $b_1$, determined by the approximate solution to Eq.~\eqref{bolt},  are given by
\ben
\left\{\begin{array}{c} b_0(a,T)\\b_1(a,T)\end{array}\right\}=\frac{3nl(T)}{C_{00}C_{11}-C_{01}^2}\left\{\begin{array}{c}-C_{01}(a,T)\\C_{00}(a,T)\end{array}\right\}\,,
\label{bs}
\een
 where 
 \ben
 l(T)={35\over4}{f_{7/2}(z)\over f_{3/2}(z)}-\left({w(T)\over k_BT}\right)^2\,,
 \label{lTeq}
 \een
$z=e^{\mu/k_BT}$ is the fugacity, $f_n(z)=-{\rm Li}_n(-z)$, Li$_n(z)$ are the polylogarithmic functions,
and $C_{nm}$ are the $2\times2$ matrix elements of the collision integral,
\begin{align}
C_{nm}&={1\over k_BT}\int{d\mbp_1d\mbp_2\over(2\pi\hbar)^6} {|\mbp_1-\mbp_2|\over m}f^0_{\mbp_1}f^0_{\mbp_2}\nn
&\times{1\over 4}\int d\Omega_r(1-f^0_{\mbp_3})(1-f^0_{\mbp_4})\nn
&\times\left\{\frac{d\s_{++}}{d\Omega_r}\D_{++}[\e_\mbp/k_BT)^n\mbp]\cdot\D_{++}[(\e_\mbp/k_BT)^m\mbp]\right.\,\nn
&\left.+\frac{d\s_{+-}}{d\Omega_r}\D_{+-}[\e_\mbp/k_BT)^n\mbp]\cdot\D_{+-}[(\e_\mbp/k_BT)^m\mbp]\right\}\,,
\label{cnm}
\end{align}
where we define $\D_{+-}[\phi_\mbp]=\phi_{\mbp_3}+\phi_{\mbp_4}-\phi_{\mbp_1}-\phi_{\mbp_2}$ and $\D_{+-}[\phi_\mbp]=\phi_{\mbp_3}-\phi_{\mbp_4}-\phi_{\mbp_1}+\phi_{\mbp_2}$ for an arbitrary function $\phi_\mbp$, and in the integrand momentum conservation is satisfied: $\mbp_1+\mbp_2=\mbp_3+\mbp_4$.  

\begin{figure}[t]
\begin{center}
\begin{tabular}{cc}
\includegraphics[width=.49\linewidth]{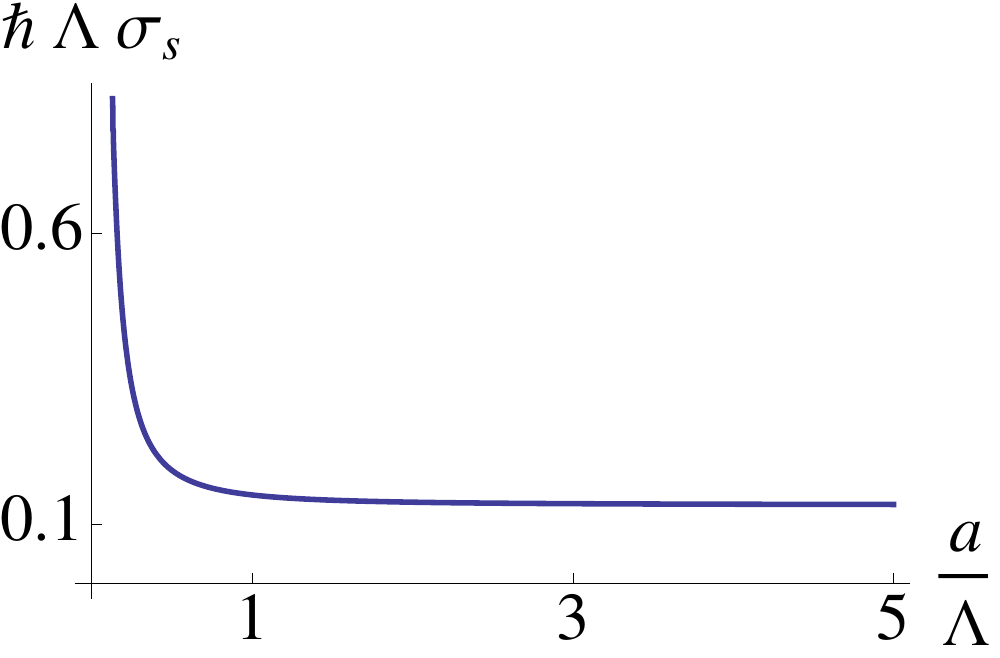}
&\includegraphics[width=.49\linewidth]{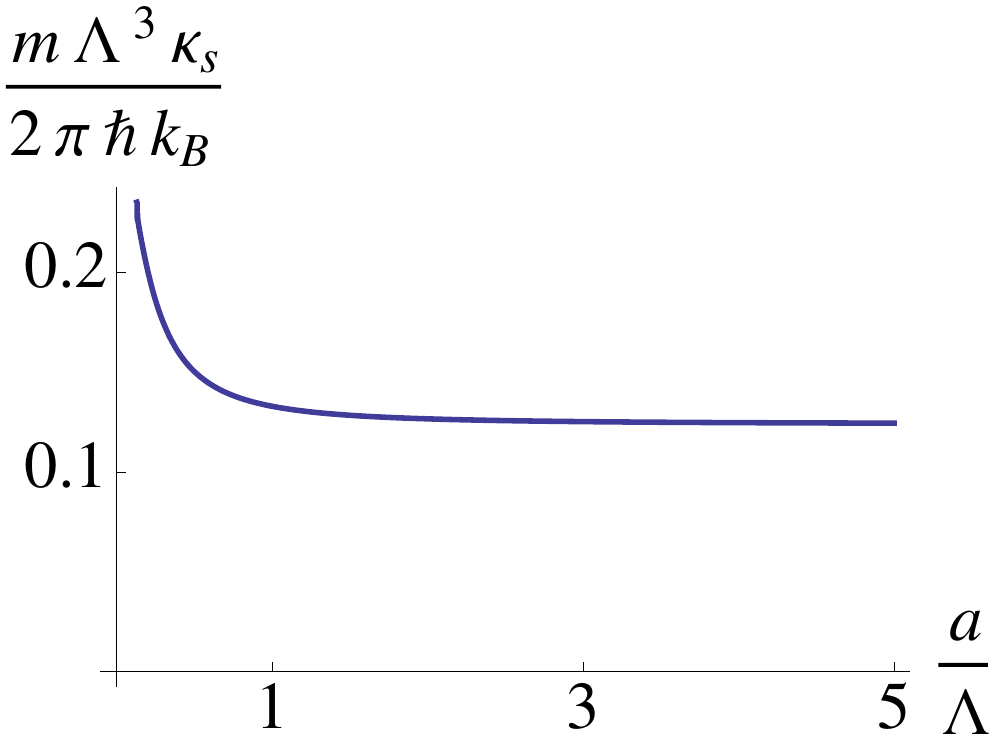}\\
(a)&(b)\\
\includegraphics[width=.49\linewidth]{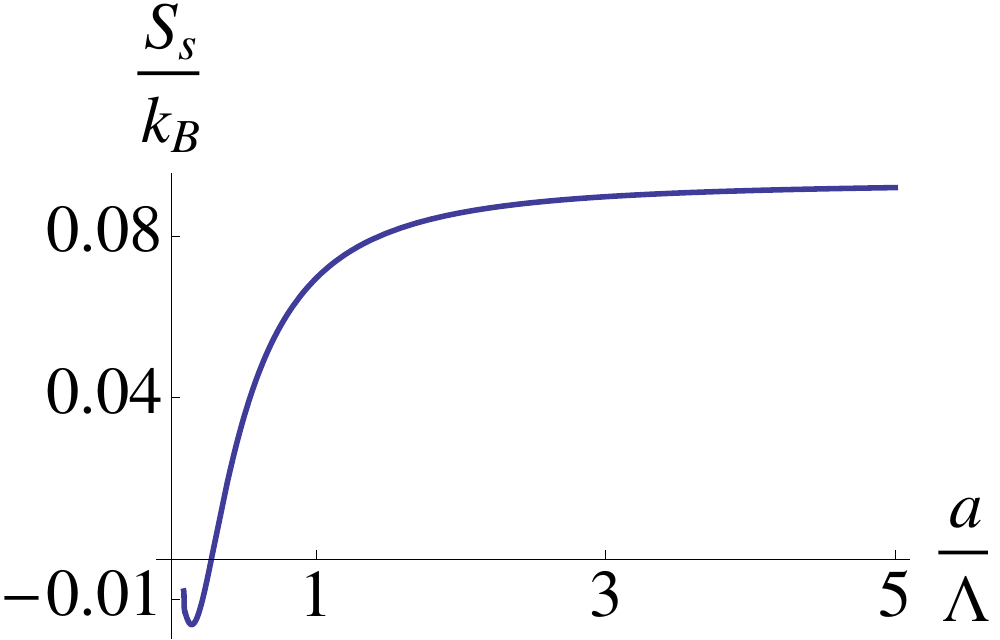}
&\includegraphics[width=.49\linewidth]{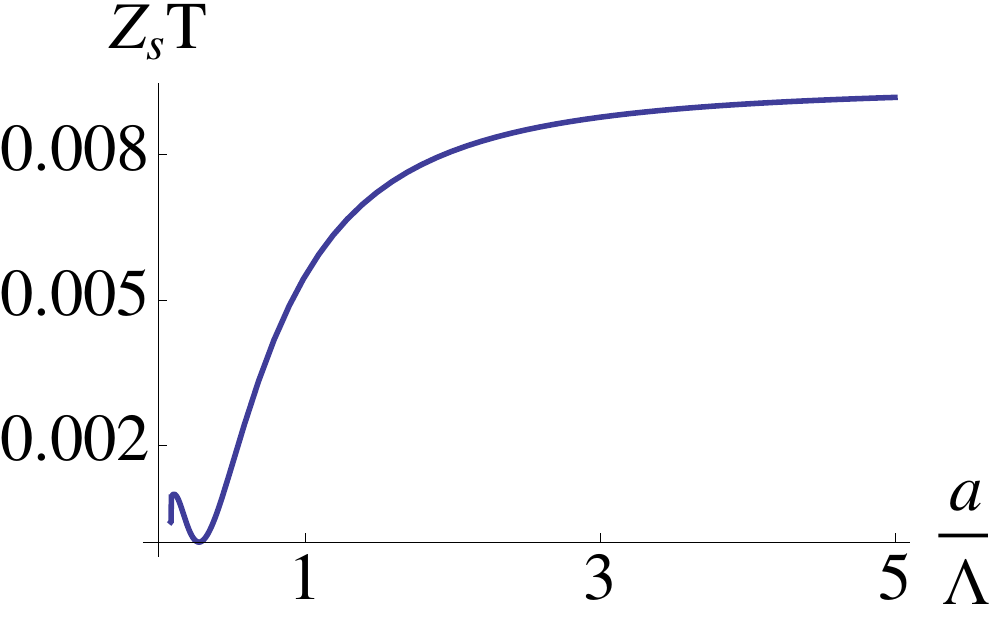}\\
(c)&(d)
\end{tabular}
\caption{Plots of the transport coefficients in the high-temperature limit, $T\gg T_F$, as a function of $a/\Lm$.  (a) The spin conductivity normalized as $\hbar \Lm\s_s$, (b) the spin heat conductivity normalized as $m \Lm^3 \kappa_s/2\pi\hbar k_B$, (c) $S_s/k_B$ and (d) $Z_sT$.
}
\label{seebeckhighT}
\end{center}
\end{figure}
From Eq. \eqref{current} and Eq. \eqref{response}, it follows that the Seebeck coefficient is given in terms of $b_0,b_1$ and the spin conductivity $\s_s$ by
\ben
 S_s(a,T)={n\over\s_s}\left[b_0(a,T)+b_1(a,T){w(T)\over k_BT}\right].
 \label{seebeck}
 \een
 The other transport coefficients are calculated similarly.  In order to make the numerics more tractable, we have omitted the angular dependence in $d\s_{++}/d\Omega$.  On the other hand, in the high-temperature limit, the integrals in Eq.~\eqref{cnm} including the angular factor can be done analytically [\onlinecite{cnm}], which allows us to calculate the coefficients in the high-temperature limit.  The result is shown in Fig.~\ref{seebeckhighT} plotted as a function of $a/\Lm$, where $\Lm=\sqrt{2\pi\hbar^2/mk_BT}$.   In this limit, the temperature dependence of $\s_s$ can be understood from classical considerations.  The spin conductivity is approximately related to the inter-spin collision time $\tau_{+-}$ by $\s_s\propto n\tau_{+-}$ and  $1/\tau_{+-}=n\bar{\s}_{+-} v_T$, where $\bar{\s}_{+-}$ is the inter-spin cross section and $v_T\propto 1/\Lm$ is the average thermal velocity.  In the limit $a/\Lm\to0$,  $\bar{\s}_{+-}\propto a^2$, thus $\Lm\s_s\propto(\Lm/a)^2$, and when $a/\Lm\to\infty$,  $\bar{\s}_{+-}\propto \Lm^2$ thus $\Lm\s_s\propto1 $, in agreement with our result.

\begin{figure}[t]
\begin{center}
\begin{tabular}{cc}
\includegraphics[width=0.5\linewidth]{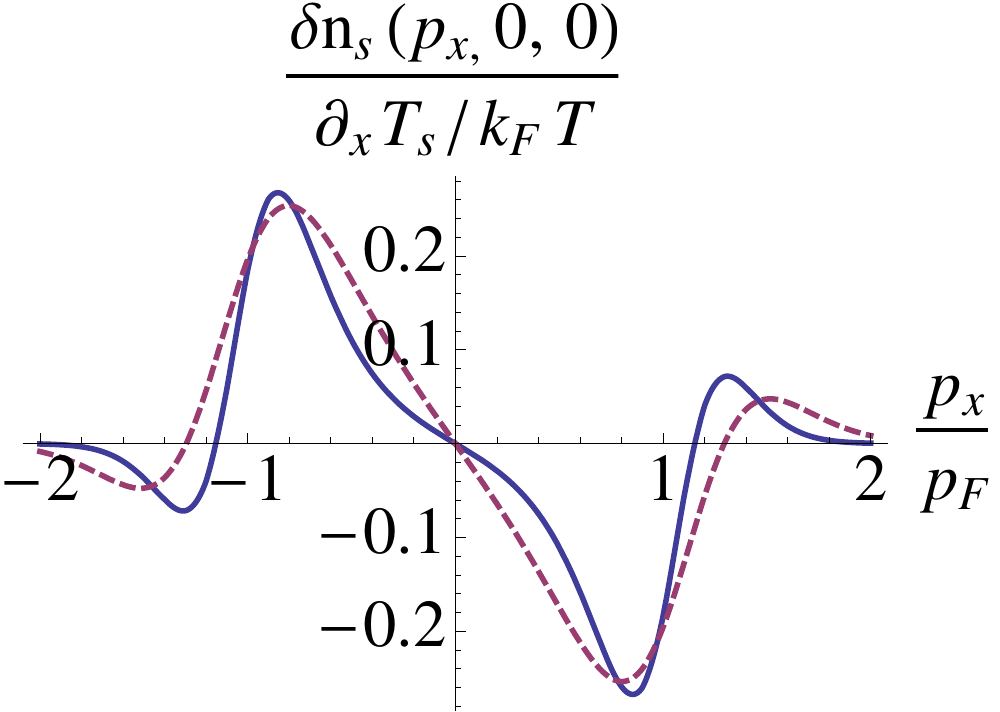}&
\includegraphics[width=0.5\linewidth]{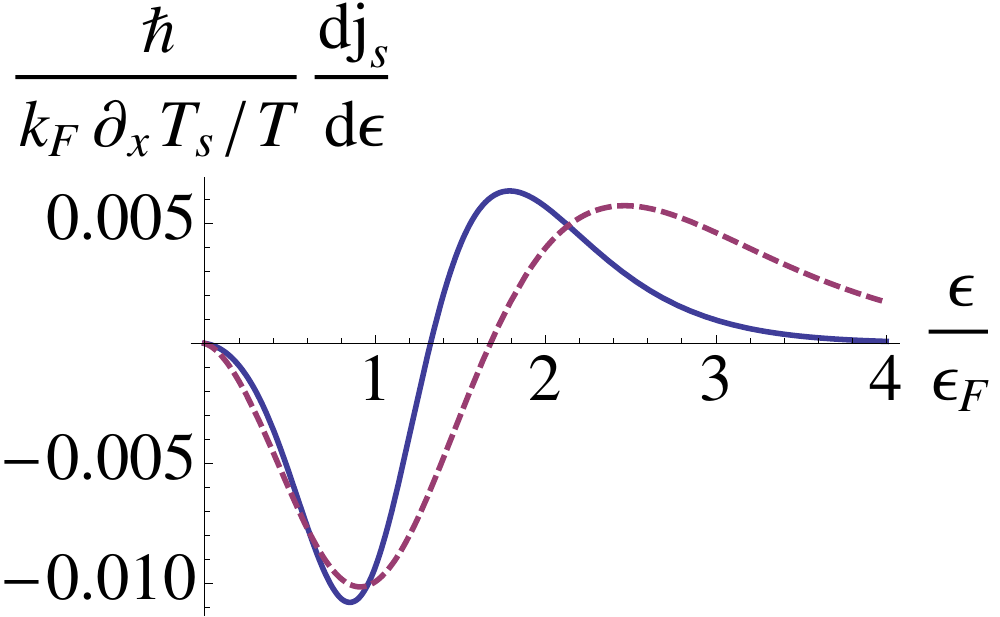}\\
(a)&(b)
\end{tabular}
\caption{ A plot of (a) the perturbed spin distribution per spin temperature gradient, normalized as $\de n_s/(-\partial _xT/k_FT)$ along the $p_x$-axis, and a plot of (b) the spin current density along the $p_x$-axis, per $d\e$ per spin temperature gradient, normalized as $(dj_s/d\e)(\hbar/k_F(\partial _xT_s/T))$,  for $k_Fa = 10$ and temperatures at which $S_s$ is negative ($T=.3T_F$, blue thick line) and positive ($T=.5T_F$, purple dashed line).}
\label{dnkfa10}
\end{center}
\end{figure}

The behavior of the transport coefficients depends crucially on the shape of the perturbed spin distribution $\de n_{\mbp s}=-\p_\e f^0_\mbp \phi_{\mbp s}$ and the associated spin current density, which we plot for $k_Fa=10$ in Fig.~\ref{dnkfa10}.  The positive (negative) parts of $\de n_{\mbps}$ may be regarded as particle (holes) having group velocities $\pm \mbp/m$.  Since $\de n_{\mbps}=-\de n_{\mbp-\s}$,  every spin up particle is matched with a spin down hole with the same momentum, resulting in the spin current. Thus, the sign of $S_s$ is determined by the relative number of particles or holes induced in response to the spin temperature gradient.


Next, we give a criterion that determines the sign of the spin-Seebeck coefficient and show that it does not depend on the form of the intra-spin scattering at all. First, we note that in our solution given by Eq.~\eqref{bs}, we always have $b_0<0$ and $b_1>0$ because $l(T)>0$,  $C_{nm}>0$, and $\det\hat{C}>0$ [\onlinecite{eigen}], which implies $b_0<0$ and $b_1>0$.  Thus for positive momenta, $b_1$($b_0$) corresponds to particles (holes) created above (below) the Fermi surface.   Inspecting Eq. \eqref{seebeck}, we find that the criteria to have $S_s\leq0$ is  
  \ben
 \left|{b_0\over b_1}\right|=\left|{C_{01}\over C_{00}}\right|\geq{w\over k_BT}\,,
 \label{criteria}
 \een
and the opposite inequality for $S_s>0$.   Thus $S_s$ is positive when $b_1$ becomes large enough to violate Eq.~\eqref{criteria} [\onlinecite{relax}].  Furthermore, it turns out that the only integral in Eq. \eqref{cnm} containing intra-spin scattering that is non-vanishing is $C_{11}$, which does not enter in Eq. \eqref{criteria}.

The temperature dependence of $b_0$, $b_1$ follows from the temperature dependence of the collision matrix elements in Eq. \eqref{cnm}, which are given by integrals nonvanishing only for ${p}_r\sim\sqrt{4\pi}\hbar/\Lm$.  Therefore, it is useful to express the differential cross section Eq. \eqref{cross} in terms of the rescaled momentum $\til{p}_r=(\Lm/\sqrt{4\pi}\hbar) p_r$,
\ben
\frac{{d\sigma}_{+-}}{{d\Omega}}={1\over k_F^2}\frac{(k_Fa)^2}{1+(k_Fa)^2(T/T_F)\til{p}_r ^2}\,,
\label{cross1}
\een
From Eq. \eqref{cross1} we see that the change in the sign of $S_s$ is related to the crossover from hard-sphere scattering, $d\s/d\Omega\simeq a^2$  when $k_Fa\ll1$ or $T/T_F\ll1$, to  momentum-dependent scattering, $d\s/d\Omega\sim \til{p_r}^{-2}$  when $k_Fa\simeq1$ and $T\simeq T_F$.

\textit{Discussions and outlook}.-- We note that as the temperature is lowered, one expects to enter the Fermi-liquid regime, for $T_c<T\ll T_F$,  where $T_c$ is temperature for the superfluid transition, and one should use Fermi-liquid scattering amplitudes in the collision integral [\onlinecite{bruunNJP11}]. Near and above $T_c$, one should also take into account effects of pairing correlations on the inter-spin interaction [\onlinecite{bruunPRA05}].  Both these regimes can be analyzed with the formalism presented in this paper, and will be relegated to future work.

This work was supported by Stichting voor Fundamenteel Onderzoek der Materie (FOM), the Netherlands Organization for Scientific
Research (NWO), by the European Research Council (ERC) under the Seventh Framework Program (FP7).


\begin{thebibliography}{15}
\bibitem{trapavg}
{{We note that the measured values of the transport coefficients should be compared with the trap-averaged values, which differs slightly from the results presented here, but may readily be computed using our Boltzmann formalism.}}
\bibitem{enthalpy}
Since $s\sim T/T_F$, for degenerate fermions, one usually takes $w\to\mu$.
\bibitem{cnm}
{Specifically, $C_{nm}$ can be expressed in terms of incomplete Gamma functions.}
\bibitem{eigen}
{The eigenvalues of the collision matrix are negative because they correspond to relaxation times, and therefore, for the $2\times2$ collision matrix, the determinant must be positive.}
\bibitem{relax}
{Note that the Seebeck coefficient vanishes in a simple relaxation time approximation given by ${\phi}_{\mbp s}=\tau_s(w-\e_\mbp)v_x\p_x T_s/T$, corresponding to the case $|b_0/b_1|=w/k_BT$ in Eq. \eqref{criteria}.}
\end{thebibliography}
\end{document}